\documentclass[]{spie}  

\usepackage[]{graphicx} \usepackage{psfig} \title{Observing the
Seyfert 2 nucleus of NGC 1068 with the VLT Interferometer}

\author{Huub R\"ottgering\supit{a},  Walter Jaffe\supit{a}, 
Klaus Meisenheimer\supit{b},
Helen Sol\supit{c}, Christoph Leinert\supit{b},
Andrea Richichi\supit{d}  and Markus Wittkowski\supit{d}
\skiplinehalf
\supit{a} Leiden Observatory, P.O. Box 9513, 2300 RA Leiden, The Netherlands; \\
\supit{b} Max-Planck-Institut f\"ur Astronomie, K\"onigstuhl 17, D-69117 Heidelberg,Germany; \\
\supit{c} Observatoire de Paris-Meudon, 5 place J. Janssen, 92195 Meudon Cedex, France; \\
\supit{d} European Southern Observatory, Karl-Schwarzschild-Str. 2, D-85748 Garching bei M\"unchen, Germany. 
}

\def\aj{AJ}                   
\def\araa{ARA\&A}             
\def\apj{ApJ}                 
\def\apjl{ApJ}                
\def\ao{Appl.~Opt.}           
\def\aap{A\&A}                
\def\mnras{MNRAS}             
\def\pasp{PASP}               
\def\nat{Nature}              

\def\spose#1{\hbox to 0pt{#1\hss}}
\def\simlt{\mathrel{\spose{\lower 3pt\hbox{$\mathchar"218$}}
     \raise 2.0pt\hbox{$\mathchar"13C$}}}
\def\simgt{\mathrel{\spose{\lower 3pt\hbox{$\mathchar"218$}}
     \raise 2.0pt\hbox{$\mathchar"13E$}}}
\def\simpropto{\mathrel{\spose{\lower 3pt\hbox{$\mathchar"218$}}
     \raise 2.0pt\hbox{$\propto$}}}

\begin{document} 
  \maketitle 

\begin{abstract}
Dusty tori have been suggested to play a crucial role in determining
the physical characteristics of active galactic nuclei (AGN), but
investigation of their properties has stalled for lack of high
resolution mid-IR imaging. Recently, a long-awaited breakthrough in
this field was achieved: NGC 1068, a nearby AGN, was the first
extragalactic object to be observed with a mid-IR interferometer, thereby
obtaining the needed angular resolution to study the alleged torus.
The instrument used was MIDI mounted on the ESO's VLT
interferometer. The resulting $8-13 \mu$m interferometric spectra
indicated the presence of a thick ($3\times 4$ parsec) configuration
of warm dust surrounding a hot $ \sim 1 $pc component, marginally
elongated in the direction perpendicular to the main orientation of the
warm component.  The structure of the 10 micron ``silicate''
absorption feature hinted at the presence of non-typical dust.

In this proceeding, first the field of AGN research is briefly reviewed,
with an emphasis on models of dusty tori. Second, the general
properties of the key object NGC 1068 are discussed. Third, the MIDI
data set is presented together with a first attempt to interpret this
data in the context of tori models. Fourth, preliminary MIDI
interferometric spectra of the nucleus of the nearby starbursting
galaxy Circinus are presented. The apparent observed absence of both a
hot component as well as a sharp absorption feature suggest that we
view the torus more edge-on than is the case for NGC 1068.  Finally,
we briefly discuss the prospects of ESA's Darwin mission for observing
nearby and distant AGN. The required capabilities for Darwin's first
goal -- the search for and subsequent characterization of earth-like
planets orbiting nearby stars -- are such that for its second goal --
high resolution astrophysical imaging -- the sensitivity will be
similar to JWST and the angular resolution 1-2 orders better. This
will allow detailed mapping of tori of low luminosity AGN such as NGC
1068 up to redshifts of 1 - 2 and more luminous AGN up to redshift of
10 and beyond.

\end{abstract}



\section{INTRODUCTION}

Active galactic nuclei reside in centers of galaxies and emit
radiation over the entire electromagnetic spectrum, from the most
energetic gamma-ray photons up to the longest radio waves. For a
significant fraction of the AGN, the emission is so luminous that it
outshines the entire associated galaxy. An intriguing fact is that AGN
activity was much more common in the early Universe than it is today:
there were a factor of 100 - 1000 more powerful AGN at $z\sim 1$ per
co-moving volume than there are in our local universe. This
suggests an intricate relation between the formation of galaxies and
the activity of AGN. The hope is that a detailed study of the
physics of AGN will lead to a deep understanding of this link.

The title of this conference is ``advances in stellar
interferometry''. Contrary to what is suggested in this title, one of
the main goals of the present rapid development in this field is to
move to a situation where optical
interferometry is a general tool in the study of many astrophysical
phenomena. The VLT observatory with its 4 large telescopes has been
especially designed with this goal in mind.  An important step in
making optical interferometry accessible for general astrophysical
usage was the interferometric 8$ - 13$ $\mu$m observations of the famous
nearby Seyfert 2 galaxy NGC 1068, using the MIDI instrument mounted on
the VLT Interferometer (VLTI).  The results from these observations
are reported in Jaffe  et 
al  (2004).\nocite{jaf04} Furthermore, the VINCI test camera
operating in the K-band and also mounted on the VLTI turned out to have
sufficient sensitivity to obtain fringes on NGC 1068 (see Wittkowski
et al 2004). \nocite{wit04} Finally, Swain et al (2003) have used the
KECK interferometer to obtain visibility measurements on a northern
Seyfert galaxy, NGC 4151. \nocite{swa03} Swain reports on these
results in this conference.

In this contribution, we will first briefly discuss some of the issues
in the field of AGN, with an emphasis on the current status of models
of dusty tori.  Second, many different kinds of observational results
on NGC 1068 are briefly touched upon.  Third, the VLTI observations on
NGC 1068 are presented and subsequently discussed in the context of
models of dusty structures in the centers of active galaxies. Fourth,
we briefly present preliminary results on MIDI observations of
the Circinus galaxy, indicating that there may be a significant
variety of structure of the dusty components in the hearts of
AGN. Finally, we briefly discuss the potential of observations of
distant AGN with ESA's space interferometer Darwin.

\section{ACTIVE GALACTIC NUCLEI} 

With more and more observational techniques becoming available in the
sixties and seventies of the last century, more and more classes of AGN
were defined.  This resulted in what is commonly referred to as the
``zoo of AGN''.  This zoo contains may different types of objects
including Seyfert 1 and 2, quasars, radio galaxies, emission line
galaxies, BLACs, etc (e.g. Antonucci 1993, Urry \& Padovani
1995). \nocite{ant93,urr95} An important aim of research into these
objects is to understand the underlying physical processes that
interrelate the various classes of AGN. In such studies a number of
very important building blocks have either been found or suggested to
be present. First, there is overwhelming evidence that the centers of
AGN harbor massive black holes with masses ranging from 10$^6$ up to
10$^9$ solar masses.  X-ray luminous accretion disks surrounds these
black holes. Part of the material that is swirling in through such
disks leaves this region though a collimated outflow, the jets. It has
been suggested that the accretion disk is surrounded by a dusty
structure, possibly in the form of a  bagel: the dusty torus. Furthermore, 
characteristic components are two different types of emission line
regions: the broad emission line regions with characteristic
velocities of 10,000 km s$^{-1}$  are thought to originate on spatial scales
similar to the accretion disk ($\simlt $ pc), while the narrow emission line
region with velocities a factor $10-50$ smaller  comes from  scales
encompassing that of the torus (few - 100 pc).

With all these building blocks a start can be made to attempt to
relate the different classes of AGN. The mass of the central black hole
has been suggested to be directly related to the maximum total energy output
of the AGN, with the most luminous AGN harboring the most massive
black holes. The amount of spin of the black hole possibly determines
whether a well formed jet will emerge. A likely way to start the
AGN activity seems to be the merging of two galaxies, which provides the
material for the various AGN building blocks. This naturally leads to
a time sequence from a building up phase, to a steady AGN, and finally 
the end of the activity. A very important role is thought to
be played by the torus. Viewed edge-on, the line of sight
towards the center is blocked, obscuring the nucleus and the broad
line region. Viewed along the poles, the very luminous core and the
broad line region are both apparent.

\section{MODELS OF DUSTY TORI} 

With the existence  of tori being an attractive way of understanding a
number of observational characteristics of AGN, efforts were started to
model the physical characteristics of the tori.  Radiative transfer
calculations were carried out for  toroidal dusty structures
surrounding an UV emitting source. The resulting IR spectrum can then be
compared to what is observed. The first calculations were carried out
by Pier and Krolik in 1992, \nocite{pie92} followed by Efthathiou and
Rowin-Robinson (1995) \nocite{efs95} and Granato et
al (1997). \nocite{gra97} In general, a good match was found for the
observed IR spectrum.  Unfortunately no strong constraints could be placed on
the general size and shape of the torus. However, at this time, this was not a major issue, since angular sizes of the tori were very much
smaller than could be observed.  Although this seemed to be a nice
and consistent scenario, a number of unresolved problems
remained. This was recently emphasized by Elitzur et al (2003) \nocite{eli03}
and the problems include: 

\begin{enumerate} 
\item X-ray observations show large
variations in X-ray column densities from object to object that are too large
to be accounted for by simple uniform torus models.

\item  Emission from
normal dust in galactic or stellar environments often shows strong
emission or absorption from the 10 micron silicate feature. The models
need a considerable amount of fine-tuning to explain both the very
modest features in Seyfert 2's and their absence in Seyfert 1's.

\item 
Some Seyfert galaxies have changed type, a behavior clearly not
expected from the uniform torus models.

\item  The stability of a
uniform torus is an important issue: collisions within such a
structure will quick lead to a disk that is substantially thinner than
is required to explain the relative fraction of Seyfert 1's to Seyfert
2's. 
\end{enumerate} 

As Krolik and Pier first noted and was recently elaborated on
in Nenkova et al 2002, \nocite{nen02} a number of these problems will
disappear if a clumpy torus is introduced. This, however, significantly
increases the number of parameters with which an obscuring torus can
be described.  It is therefore clear that the best way to establish the
morphology of tori is to use high angular resolution observations.

\section{NGC 1068} 

A prime target in the field of high resolution observation of the centers
of AGN is NGC 1068. It is the closest and brightest in the IR of all
AGN. With over 200 refereed papers containing its name in the title, it
is the most studied AGN.  Its nearly face-on orientation gives a
beautiful view on its intricate spiral structure. A good point to
start a discussion of the role of NGC 1068 in the understanding AGN
physics is the important observation by Antonucci and
Miller (1985). \nocite{ant85} They showed that the polarized light of this narrow
line object bears the signature of a broad line Seyfert 1 object. This
naturally led to the conjecture of the presence of tori.  The HST
imaging of both narrow line and UV emission showed clear conical
regions (Macchetto et al 1994). \nocite{mac94} This was also
interpreted in a scenario in which a torus permits only radiation from
the nucleus to penetrate in two preferred and opposite directions.
A next important high light was the observations of H$_2$0 maser
emission (Greenhill et al 1996). \nocite{gre96} Such emission
originate in fairly warm (400 K) and dense ($10^8 - 10^{10}$
cm$^{-3}$) gas.  In a velocity - position diagram, the measurements
can be fitted by gas in a simple Keplerian orbit around a black hole. A
more elaborate model by Lodato and Bertin (2003) \nocite{lod03}
suggests that the observed disk is self gravitating. They deduce 
for both  the mass of the black hole and the disk $8 \times 10^6$ M$_0$.  Sensitive
VLBA measurements showed a pc-scale size disk interpreted as due to
free-free emission from an ionized dense (10$^6$ cm$^{-3}$) and hot
($10^7$ K) gas (Gallimore et al 1997). \nocite{gal97} Schinnerer et al
2000 \nocite{sch00} concluded from their detailed mapping of the 12
CO(2--1) line that there was evidence for a warped 100 pc CO emitting
disk enclosing a mass of 10$^8$ M$_0$. Finally, Chandra images
revealed a detailed correspondence between the optical and X-ray
morphologies, also indicating that the extended X-ray emission is
beamed nuclear emission (Ogle et al 2003). \nocite{ogl03} The
spectrum of the nucleus seems to be best fit by a power-law obscured by
a Compton structure, possibly related to the torus (Matt et
al 1997). \nocite{mat97} Finally, deep images in the near and mid
infrared have been taken by a number of authors (see for example
Wittkowski et al 1998; Bock et al 2000; Gratadour et al 2003; Rouan
et al 2004 and references therein). \nocite{wit98,boc00,gra03,rou04} These
images show a compact core with a number of addition morphological
features within a region of 1 arcsec from this core. Most notably is
the ``tongue'', a northern extension from the nucleus oriented 
in the same direction as the radio jet.

\section{VLT interferometer} 

Currently the VLT interferometer can make use of the four 8.2 meter VLT
Unit Telescopes (UTs). Ultimately, when the 1.8 m Auxilary Telescopes (ATs) are
installed the maximum attainable baseline is 2005 m. 
A test camera by the name of  VINCI can combine beams from two UTs
and detect fringes in the K-band. The first science instrument, MIDI,
currently observes in the wavelength region 8 -13 micron. For a
detailed description we refer to Leinert et al (2003). \nocite{lei03}
An update to the 18-22 micron region is expected to become available
in 2006. Finally, AMBER (see Petrov et al 2003) \nocite{pet03} is also
installed and operational at the VLTI and will be offered for general
usage in 2005.  This instrument will allow for simultaneous
observations in the  J, H, and K bands yielding in J-band a resolution of 1
mas for the maximum baseline of 205m.  The sensitivity of AMBER and
MIDI are such that tens of AGN should be observable. As discussed,
torus models are not accurate enough to reliably predict the amplitude
of the visibilities.  Hence, there was the possibility that tori would
be extended and smooth with low visibilities that would make it
impossible to interferometrically observe nearby AGN with the first
phase of VLTI instrumentation.  

The potential of the VLTI to deliver high angular studies
of AGNs will be further increased by two systems, FINITO
and PRIMA. The former is already installed and under testing,
the latter is to be available in 2006. FINITO is the external fringe tracker,
which
will permit integration times much longer than the atmospheric
coherence time (hundredths of a second).
PRIMA will permit differential phase measurements
between a scientific target and a nearby reference star, achieving
true imaging and astrometry at the ten-microarcsecond level.
The resulting drastic increase in sensitivity would then
certainly bring many AGN in the realm of IR interferometric observing.

\section{MIDI observations of NGC 1068} 

As discussed, one of the brightest nearby AGN is NGC 1068.  Our
observations of this object were obtained during MIDI's initial
scientific runs in June and November, 2003 and reported in Jaffe 
et al 2004.  Figure 1a shows the
corresponding N-band spectrum of the source using the grism 
with a resolution R=30, as integrated over
the beam width of 0.4 arcsec.  Fig. 1b and 1c show the spectra as seen
by the two-telescope VLT interferometer operating at the projected
baselines: $B = 42$ m (resolution $\sim 50$ mas, orientated at
a position angle $P.A. \simeq 40\deg$ East of North), and $B = 78$ m
(resolution $\sim 26$ mas, $P.A. = 2\deg$ i.e. almost the direction of 
the radio jet) respectively.

\begin{figure}[h!]
\centerline{\psfig{file=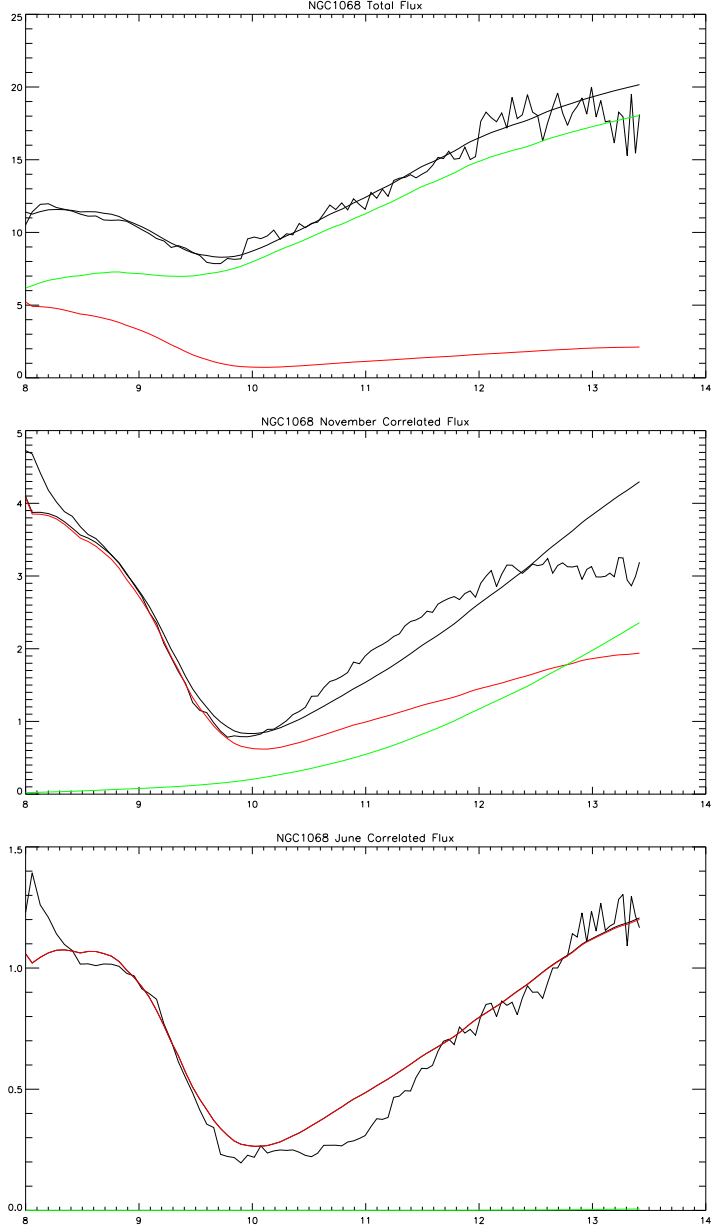,width=100mm,angle=0}}
\caption{\label{fig}
 {\bf a.} The single telescope,
 non-interferometric MIDI flux-density spectrum of the nucleus of NGC~
 1068.  The jagged line shows the MIDI data and the smooth line the
 best fitting two-component Gaussian model described in the text.
 The upper dashed line shows the contribution of the warm component
 and the lower dashed line that of the hot component.
 The absorption dip near 10~$\mu$m is caused primarily
 by `astronomically common', olivine-type, silicate dust.
 {\bf b.} The two-telescope interferometric spectrum from November 2003
 with an effective resolution of $\sim 50$ mas.
 {\bf c.} The interferometric spectrum from June 2003 with a resolution
 of $\sim 26$ mas.  Here the dip near 10 $\mu$m is probably
 caused by dust of a an  lumino-silicate or other, non-olivine, composition.
(taken from Jaffe et al 2004).
}
\end{figure}

The simplest model, which well accounts for the spectra of both the
total and interferometric flux in our measurements, requires only
{\it two} Gaussian components of different temperature, size and
foreground dust absorption $\tau_{SiO}$.  The first is a {\it hot}
($T > 800$~K), compact component for which we fix the temperature
to nominally 1000~K.  Its length in the direction of the radio jet is well
constrained to 1.1 pc; along the orthogonal direction we can set
an upper limit $< 0.9$ pc.  The second is a {\it warm},
well-resolved component ($3 \times 4$ pc) with a best-fit
temperature $T = 270$~K. The 10~$\mu$m absorption feature in the
highest resolved interferometric observations (Fig.\ 1c) does not
fit well to the profiles of common olivine-type silicate dust
because those begin to drop already at $\sim 8\mu$m, while the
interferometric spectrum does not seem affected shortward of 9
$\mu$m. Even if the observed profile might be modified by
radiative transfer effects, we obtain a much better fit by using
the profile of calcium aluminum-silicate (Ca$_2$Al$_2$SiO$_7$), a
high temperature dust species found in some supergiant stars.


This simple heuristic model describes the basic observed scale
sizes of the dust emission in NGC~1068. 
The torus models as described in section 3 
can be divided in two classes: {\it Compact}
tori (Pier and Krolik 1995), where the dust fills a thick cylinder
(with an axial hole) of a few parsecs in diameter, and {\it extended}
fat disk models (Efstathiou and Rowin-Robinson 1995; Granato et
al 1997), with tens to hundreds of parsecs in  diameter.  Common to
both classes was a very high absorption optical depth ($A_V >
100$, i.e. $\tau_{12\mu} >4$) in the equatorial plane. This leads
to the prediction that hot dust can only be observed from those
surfaces which are both freely exposed to the radiation from the
accretion disk and unobscured along the line-of-sight from the
observer.  In consequence, the observable structures at 270 K and
at $1000$ K are almost identical and suffer very similar dust
absorption. Obviously, this is not what we observe in NGC 1068.

Already the unification between Seyfert 1 and 2 galaxies
(Antonucci 1993) seems to demand a geometrically thick dust
distribution, the height $h$ (above the midplane) of which should
be similar to its radius $r$, i.e. $h/r \simeq 1$. By resolving
the dust structure in NGC\ 1068 with MIDI/VLTI, we have
demonstrated now that indeed $h/r \simgt 0.7$, even when allowing
for projection effects (the axis of NGC\ 1068 is inclined by $\simlt
20\deg $ out of the plane of sky). Moreover, this thick structure
is located at $r \le 2$ pc. Thus, irradiation by the nuclear
source will produce a hot inner wall -- a ``funnel'' which we
identify as our hot component. Since the average cloud
temperatures are low, $T < 1000$~K, there is no way that gas
pressure could support this structure against gravity in the
nuclear potential of NGC\ 1068. Turbulent motions with average
velocities $\langle v_T \rangle \simeq 100$ km\ s$^{-1}$ (i.e.
similar to the random velocities which support the nuclear star
cluster) have to prevent the cloud system from collapsing.
However, collisions between those clouds are highly inelastic: The
turbulent motion would be damped within about one orbital period,
$t_{orb}{\rm (2pc)} \simeq 10^5$ years. Thus, a continuous
injection of kinetic energy into the cloud system seems to be  required.
To our knowledge, none of the current models of active galactic
nuclei provide a convincing solution to this problem. Realistic
torus models require both a physically motivated distribution of
the dusty molecular clouds  and a full 3-dimensional treatment of
the radiative transfer. Thus, the first high resolution
observations of the nucleus of an active galaxy by infrared
interferometry  forces a reconsideration of the physics of these
spectacular objects.

\section{VINCI observations of NGC 1068}

The VINCI interferometric K-band observations of the core of NGC 1068
used a baseline of 46 meters and have been reported in Wittkowski et al
2004. \nocite{wit04}  The deduced squared visibility is 16 \%. 
The simplest model consisted
of a single Gaussian intensity distribution with a FWHM of 0.3 pc.
Taking existing Speckle observations into account, a somewhat more
complicated analysis constrained 40\% of the flux to come from scales
of clearly below 0.3 pc  and the
remainder of the flux originating from scales smaller than the FOV, being
3.3 pc. The 0.3 pc  scale is  to be rather similar to what Swain et
al found for the core of NGC 4151.  See also Swain's discussion during this
meeting. With only one visibility point, it is clear that a discussion
on the physical origin of such a compact structure is somewhat
premature. Taking the suggested results of  MIDI  at face
value, a likely possibility is that the compact VINCI component can be
directly related to the 1000 K component, as observed with MIDI.  The
VINCI K-band spectrum would then originate from a hot funnel forming the inner
boundary of the torus. Alternatively, the emission might come directly
from the outer part of the hot accretion disk. Further
observations that constrain for example the orientation of the
structure will help distinguishing between these two possibilities.

\section{The near future} 

Over the last three years we have been preparing a sample of 23
southern galaxies with a 10 micron flux density sufficiently
bright to warrant MIDI interferometric observations (central flux
densities $>400$ mJy). All these sources have been observed at 10
$\mu$m with the infrared imaging camera mounted on ESO's 3.6 m
telescopes. These observations allowed us to measure the core flux
on scales of 0.5 arcsec and obtain accurate positions. In
addition, we are in the process of obtaining diffraction limited
observations with NAOS-Conica in the HLM bands with the aim of
deriving extinction maps, identifying outflow phenomena and
determining the fraction of unresolved flux.

For the next three years, the goal of the MIDI program is two fold:

\begin{itemize}

\item An interferometric snapshot survey of all the sources from the
entire list of targets. The plan is to measure two visibility points
for each of two orthogonal baselines. This will deliver a good
indication of the size of the thermal emission regions.

\item A detailed interferometric study of a few selected targets. During
2004 and 2005 we will start this  by obtaining 
further  visibility points for NGC 1068.  This will
be used to investigate whether a simple torus-like morphology
is a correct description or whether higher spatial frequency
modifications are required (e.g., warps or clumpy tori).

\end{itemize} 

In 2005-2006,  VISIR can be used to take
detailed spectra of the  objects in this sample.
This will allow us to observe  the dust properties on scales of
0.5 arcsec, thereby addressing whether the dust properties change
dramatically with distance from the nucleus as suggested by the
NGC 1068 observations.  This data will also be used to
investigate the  influence of a circumnuclear starburst on  AGN
activity.

At this time MIDI will be augmented with a 20-micron observing
capability that we will use to map the outer, cooler regions of
the torus.  We hope to establish the connection between the warmer
structures already observed and the larger (100-1000 pc) scale
dust structures seen in many AGNs.

\begin{figure}[h!]
\centerline{\psfig{file=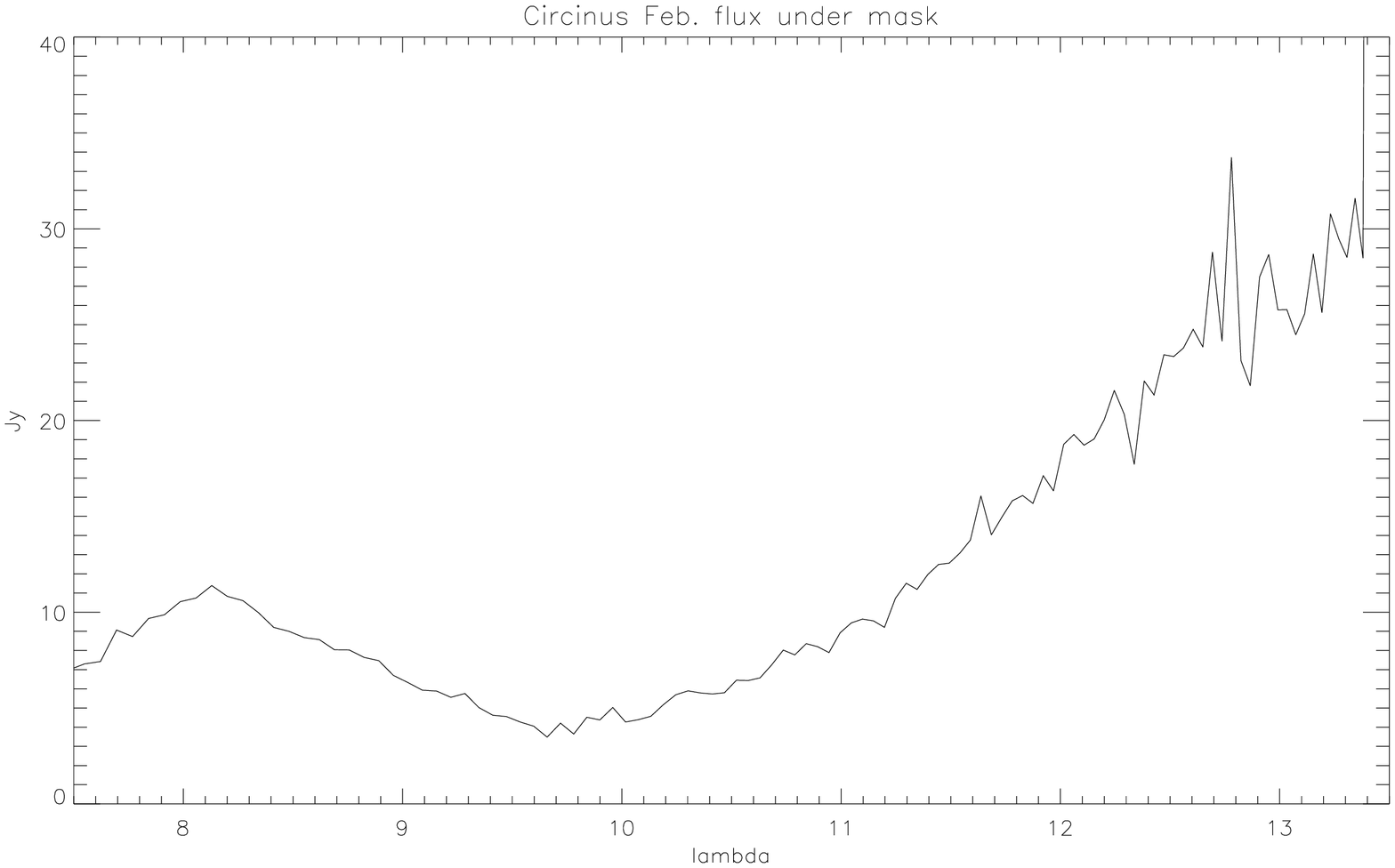,width=100mm,angle=90}}
\centerline{\psfig{file=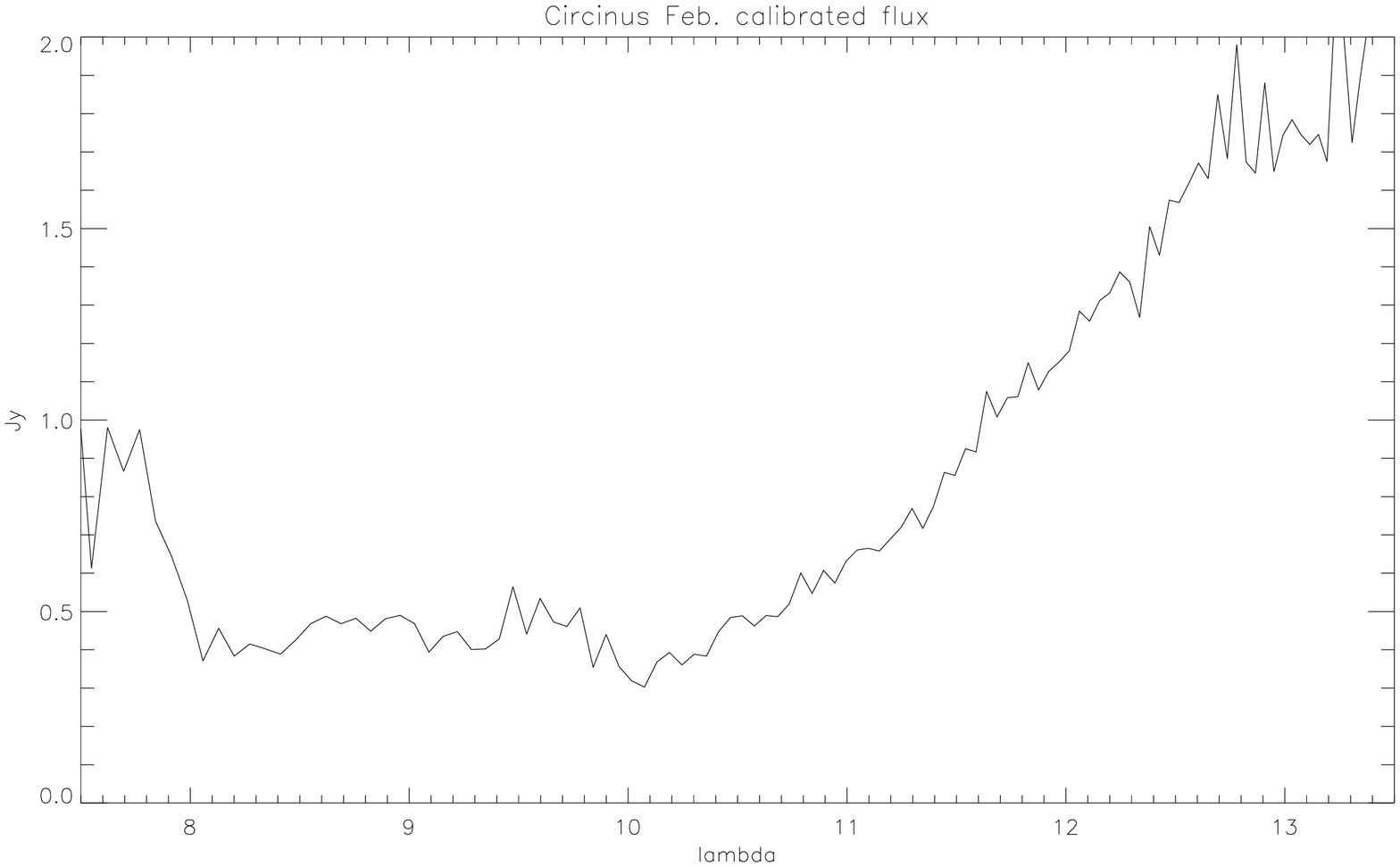,width=100mm,angle=90}}
\caption{\label{fig2}
 {\bf a.} The single telescope,
 non-interferometric MIDI flux-density spectrum of the nucleus of CIRCINUS. 
 {\bf b.} The two-telescope interferometric spectrum 
with an effective resolution of $\sim 50$ mas. }
\end{figure}

One step in this program was the observations of Circinus. This
southern Seyfert 2 galaxy with a circum nuclear starburst has many
characteristics that are distinctly different from NGC 1068.  In
Fig. \ref{fig2} we show the {\it preliminary} reduction of the total
and interferometric MIDI spectra taken in Feb 2004, using a projected
baseline of 44 m, yielding a resolution of $\sim 50$ mas. 

Comparing the interferometric spectra of Circinus to that of NGC 1068,
a number of differences are immediately apparent. First, the hot
component that is  strong at the shorter wavelengths for NGC 1068
seems virtually absent for Circinus. This might be due to orientation
effects.  If indeed the hot component in NGC 1068 is due to a compact pc
scale size structure within the torus, then such a structure could be
hidden from our view if the torus were viewed more edge-on. This
could be the case for Circinus where nuclear activity might be more
hidden and the starburst more apparent. Second, the distinctly steep
absorption, as seen in the NGC 1068 spectrum, is much less apparent for
NGC 1068. This could also be due to orientation effects, if indeed the
obscuring non-typical dust would be close to the pc scale structure
which is hidden from our view.

\section{The Future: ESA's Darwin mission} 

The proposed infrared space interferometry mission Darwin has two main
aims: (i) to detect and characterize exo-planets similar to the Earth,
and (ii) to carry out astrophysical imaging in the wavelength range $6
- 20 \mu$m at a sensitivity similar to JWST, but at an angular
resolution up to 100 times higher.  During this meeting Fridlund and
Karlsson have discussed the Darwin mission and its use to find
exo-earths. Here we will briefly discuss the performance of the Darwin
imaging mission with an emphasis on the prospects for studying AGN
(see also R\"ottgering et al 2003) \nocite{rot03a} 

The present configuration for Darwin consists of 6 telescopes each
with a diameter of 1.5 m and a central beam-combiner. Recent studies
have shown that good  performance can also be obtained in a very
cost-effective way with fewer telescopes, each with a somewhat larger
aperture. We again refer to the contributions of Fridlund and
Karlsson during this meeting for a more detailed discussion of 
the trade-off involved. 
The
relevant systems at the telescopes and the beam-combiner will be
passively cooled to 40 K, so that the sensitivity will be limited
by shot-noise from emission by the zodiacal background
(Fridlund et al 2001, Nakajima and Matshura 2001).
\nocite{fri00,nak01} 

This system should be a able to detect a
point source of 2.5 $\mu$Jy with a signal-to-noise of 5 in one hour
of integration time. Such a sensitivity is comparable to that 
which is expected to be obtained with NASA's JWST mission. 
For sources that are more complex, assessing
the integration time depends on the details of the morphology
and the observing strategy 
(see R\"ottgering et al 2003). \nocite{rot03a} 
The maximum baselines that are foreseen are about 500 meter, which
translates into a maximum resolution of 4 mas at 10 micron. This
is a factor of more than 100 higher than the nominal resolution of 350 mas at
10 micron of the 6.4 m mirror of JWST.

While the telescopes are moving it is essential that the system is
able to continuously measure the complex  visibilities.  
To be able 
to do this, the array needs to remain co-phased while moving. Since 
in general  the science targets are too faint to give enough signal
for co-phasing, the light from an off-axis bright reference
star will be  used. A system based on such a principle is currently
being implemented at the VLTI under the name PRIMA, which stands
for Phase-Referenced Imaging and Microarcsecond Astrometry (e.g. Paresce et al
2003). \nocite{par03}

\begin{figure}[htb!]
\centerline{
\psfig{figure=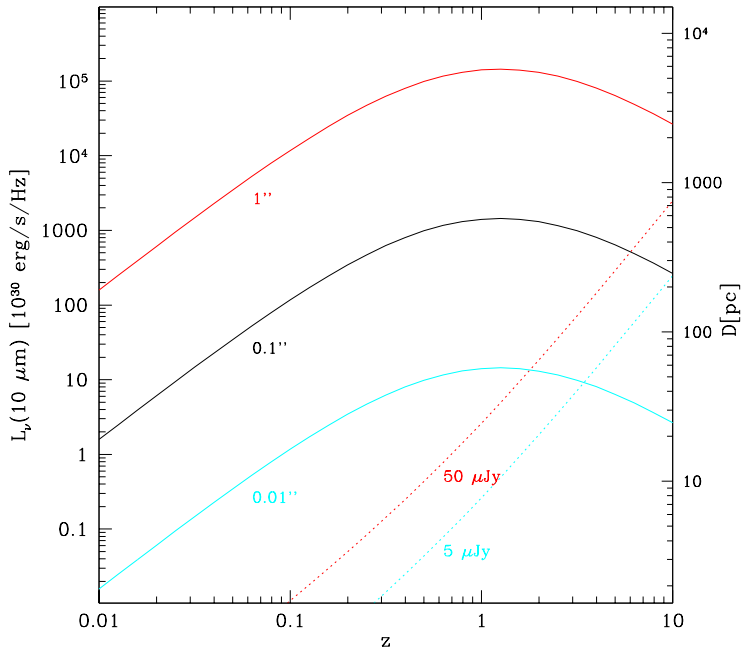,width=10cm} }
\caption{\label{agn} The 10 micron luminosity and physical scale D of
dusty tori as a function of redshift.  The solid lines
are for observed angular scales
of 1, 0.1 and 0.01 arcsec.  The dotted lines are for
an observed  10 micron flux density
of 5 and 50 $\mu$Jy.  The computations have been done for $H_0 = 75$ km
s$^{-1}$ Mpc$^{-1}$ and $\Omega=1$
(from R\"ottgering et al 2003). }
\end{figure}

With these sensitivities and resolution, it is very clear that Darwin
will be able to map nearby tori in exquisite detail.  An interesting
question is to what distance the tori of AGN can be mapped.
To investigate this, we will use the torus models of Granato et al 1997 
and results as presented in R\"ottgering et al 2003. \nocite{rot03a}  
In these
models, the inner radii of these tori, $r_{\rm in}$, are set by the
distance from the central source at which the dust grains sublimate
due to strong nuclear radiation. This radius is larger for more
luminous AGN. For the models of Granato et al, $r_{\rm in} \sim 0.5
L_{46}^{1/2}$ pc, where $L_{46}$ is the luminosity of the central
optical UV emitter in units of $10^{46}$ ergs s$^{-1}$.  As a scale
size of the torus D we will use 300 $r_{\rm in}$.  In Figure
\ref{agn}, the diameter D and the 10 micron luminosity (which are
now directly coupled) are given as a function of $z$ for angular scales of
1, 0.1 and 0.01 arcsec.  The dotted lines correspond to the 10 micron
luminosities as a function of $z$ for 10 micron flux densities of 5
and 50 $\mu$Jy.

A relatively weak AGN such as NGC 1068 has a 10 micron luminosity on
the order of $1.7 \times 10^{31}$ erg s$^{-1}$ Hz$^{-1}$ and its
modeled torus size is 60 pc. Note that the 
MIDI observations indicate that tori are likely to be 
significantly smaller. Up to redshifts of $z=1-2$ such weak AGN
are bright enough to make a map with 
a ratio of total signal over noise per angular resolution element of more than 
50. Also 
Darwin's 
resolution  is very adequate for imaging
the tori at these redshifts.  Brighter AGN can basically be mapped up
to a redshift of $z=10-20$ (if they exist).

This shows that Darwin cannot only study the physics of dusty tori in
our local universe, but also at large redshifts.  This will give the
unique opportunity to investigate how the properties of tori change
with redshift and when and how these tori and their associated massive
black holes are built up at an epoch when galaxies are forming.

\bibliographystyle{spiebib}   

 \end{document}